\documentclass[fleqn,usenatbib]{mnras}

\usepackage{newtxtext,newtxmath}

\usepackage[T1]{fontenc}

\DeclareRobustCommand{\VAN}[3]{#2}
\let\VANthebibliography\thebibliography
\def\thebibliography{\DeclareRobustCommand{\VAN}[3]{##3}\VANthebibliography}

\usepackage{graphicx}	
\usepackage{amsmath}	
\usepackage{comment}
\usepackage{siunitx}
\usepackage{multirow}
\usepackage[table]{xcolor}
\usepackage[normalem]{ulem}
\DeclareSIUnit\erg{erg} 

\definecolor{celeste}{RGB}{204,229,255}
\definecolor{celesteoscuro}{RGB}{0,170,228}
\definecolor{verdeclaro}{RGB}{220,255,220}
\definecolor{anaranjado}{RGB}{255,128,0}
\definecolor{darkgreen}{RGB}{0,100,0}

\title[Mid-Infrared Time Delay in Flare Footpoints]{First time delay observation between two mid-infrared channels in solar flare footpoints}

\author[M. Rojas-Quesada et al.]{
Miguel Rojas-Quesada,$^{1,2}$
\thanks{E-mail: m.rojas-quesada.1@research.gla.ac.uk}
\href{https://orcid.org/0000-0002-3733-1714}{\includegraphics[scale=0.07]{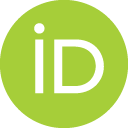}}
Lyndsay Fletcher$^{1,3}$
\href{https://orcid.org/0000-0001-9315-7899}{\includegraphics[scale=0.07]{img/ORCIDiD_icon.png}}
Hugh Hudson$^{1,4}$
\href{https://orcid.org/0000-0001-5685-1283}{\includegraphics[scale=0.07]{img/ORCIDiD_icon.png}}
Sargam M. Mulay,$^{1}$
\href{https://orcid.org/0000-0002-9242-2643}{\includegraphics[scale=0.07]{img/ORCIDiD_icon.png}}
 \newauthor{ Paulo J. A. Sim\~{o}es$^{5,1}$
 \href{https://orcid.org/0000-0002-4819-1884}{\includegraphics[scale=0.07]{img/ORCIDiD_icon.png}}}
\\
$^{1}$School of Physics \& Astronomy, University of Glasgow, Glasgow G12 8QQ, UK \\
$^2$School of Physics, Costa Rica Institute of Technology, Cartago, Costa Rica. \\
$^{3}$Rosseland Centre for Solar Physics, University of Oslo, PO Box 1029 Blindern, NO-0315 Oslo, Norway \\
$^{4}$Space Sciences Laboratory, University of California, Berkeley, CA 94720 USA\\
$^{5}$Center for Radio Astronomy and Astrophysics Mackenzie (CRAAM), Engineering School, Mackenzie Presbyterian University, S\~{a}o Paulo, Brazil
}

\date{Accepted XXX. Received YYY; in original form ZZZ}

\pubyear{\the\year{}}

\begin{document}
\label{firstpage}
\pagerange{\pageref{firstpage}--\pageref{lastpage}}
\maketitle

\begin{abstract}

The strong correlation between energy injection and mid-infrared (mid-IR) emission observed during solar flares can be used to probe energy deposition throughout the chromosphere, since the IR tracks prompt flare-induced changes in electron density.
Despite its diagnostic value, solar mid-IR observations are relatively recent, with sporadic campaigns over the last decade resulting in only a few recorded flares.
Earlier studies found time lags between mid-IR emissions from spatially resolved footpoints, offering clues about flare energy transport.
Building on this, we analyse the time lags between emissions at two wavelengths (\SI{5.2}{\micro\meter} and \SI{8.2}{\micro\meter}) for each footpoint.
Using a local cross-correlation function, we show for the first time that the \SI{8.2}{\micro\meter} emission channel peaks \SI{0.3}{\second}–\SI{0.45}{\second} before the \SI{5.2}{\micro\meter} channel. We investigate the origin of this lag, obtaining infrared emission estimates using results from the RADYN radiation hydrodynamics code. 
The theoretical lag values fall within the range of the observed ones. Variations in opacity—primarily due to flare-induced ionization—explain the wavelength-dependent temporal shift between emission maxima. In particular, longer wavelengths exhibit a smaller lag  between the peak of energy injection and peak of intensity.
These results contribute to a better understanding of how energy deposition during a flare affects the chromospheric layers of the atmosphere.
Future observations with higher temporal resolution could exploit measurements of these time lags to more fully characterize the dynamics of energy deposition during solar flares, opening a new avenue for studying heating and energy transport processes in the solar atmosphere.
\end{abstract}

\begin{keywords}

Sun: atmosphere $-$ Sun: activity $-$ Sun: chromosphere $-$ Sun: flares  $-$ Sun: infrared $-$ Sun: X-rays
\end{keywords}



\section{Introduction}
Recent observations have revealed strong mid-IR (from 5 to \SI{12}{\micro\meter} wavelength) continuum emission from solar flares, a novel view that provides valuable new insights about flare energetics. These  powerful, transient events on the Sun emit radiation from radio to $\gamma$-rays \cite{Fletcher2024}, but there are very few observations in the mid-IR regime. This is likely due to challenges of conducting solar observations at these wavelengths from the Earth’s surface.  According to \cite{Penn2016SPECTRALMID-INFRARED}, the infrared-submillimeter band can provide a vital link between the powerful photospheric emission of the white-light flare itself and the dynamics of the upper atmosphere, including coronal mass ejections and related space weather.

Previous observations of M2.0 to X9.7 flares at \SI{10}{\micro\meter} have shown that the mid-IR emission arises from compact regions and exhibits impulsive behaviour. These emissions are temporally coincident with peaks observed at microwave, visible, EUV, and hard X-ray (HXR) wavelengths \citep{Kaufmann2013, Miteva2016, Guigue2018}. Moreover, a temporal correlation between mid-IR and UV fluxes, previously reported for strong flares, was also identified in a smaller C7 event \citep{Lopez2022}. More recently, \citet{Yang2025} presented a unique observation of an X6.4 flare using the Mid-InfraRed Imager (MIRI) now installed at the Goode Solar Telescope, providing strong evidence for a chromospheric origin of the mid-IR source. Pioneer instruments operating in this wavelength range—such as AR30T, SP30T, and HATS \citep{Kaufmann2008, Kudaka2015, Guigue2020}, designed and operated by CRAAM—offer valuable opportunities to further investigate flares at mid-IR wavelengths.

The observations reported by \cite{Penn2016SPECTRALMID-INFRARED} of mid-IR emissions using MIRI during a flare suggest that these follow the non-thermal energy release of the impulsive phase. The authors have observed that the appearance of emission patches strongly suggests that they can identify these impulsive mid-IR sources with HXR flare footpoints.
Their combination of excellent time resolution, and strong, well-resolved sources make infrared observations very useful for monitoring energy deposition processes in the chromosphere during a flare. 
 
It is not yet understood how energy is transported from the corona, where it is stored, to the chromosphere, where it is radiated. To address this question, \cite{Simoes2024PreciseObservations} present a time-delay analysis of the infrared emission from two chromospheric sources in a flare. By cross-correlating the intensity signals, measured with 1 s cadence, they find a delay of $0.75 \pm 0.07 \mathrm{~s}$ at $\SI{8.2}{\micro\meter}$ and $0.73 \pm 0.16$ at $\SI{5.2}{\micro\meter}$, 
and contend that the time lag is larger than can be explained by energy transport dominated by non-thermal electrons precipitating from a single acceleration site in the corona. (We note also that these authors used the standard definition of the cross-correlation function (CCF), which tends to underestimate the lag \citep{Welsh1999OnNuclei}; we will present a less biased approach in the next section.) That result suggested energy transport by some  means other than electron beams originating in the corona, or inhibition of coronal electron energy transport.  The latter might be dominated by wave-particle interactions \citep[e.g.][]{Kontar2012Wave-particleFlares}.

Based on the same observations used by \cite{Penn2016SPECTRALMID-INFRARED}, 
and using an improved technique for measuring the time lag, we present the first observation of a delay between the two channels in the mid-IR. This delay is then interpreted in terms of the infrared emission calculated from F-CHROMA grid results \citep{Carlsson2023TheModels}, which include RADYN simulations of the evolution of solar atmospheric conditions, considering different parameters for the electron beam generated during flares.  In Section~\ref{sec:observations} we present the flare and its evolution, in Section~\ref{sec:analysis_results} we describe the analysis, the resulting measured time delays. The method of calculating IR emission is given in Section~\ref{sec:IR_radiation_flare} and Section~\ref{sec:RADYN_analysis} shows the synthesis of IR light curves from RADYN simulations. Our conclusions are presented in Section~\ref{conclusion}.

\section{Description of observations}
\label{sec:observations}

\subsection{Overview} 
\label{subsect:overview}

\begin{figure*}
    \centering
    \includegraphics[trim=0cm 1.0cm 0cm 1.0cm, width=1\textwidth]{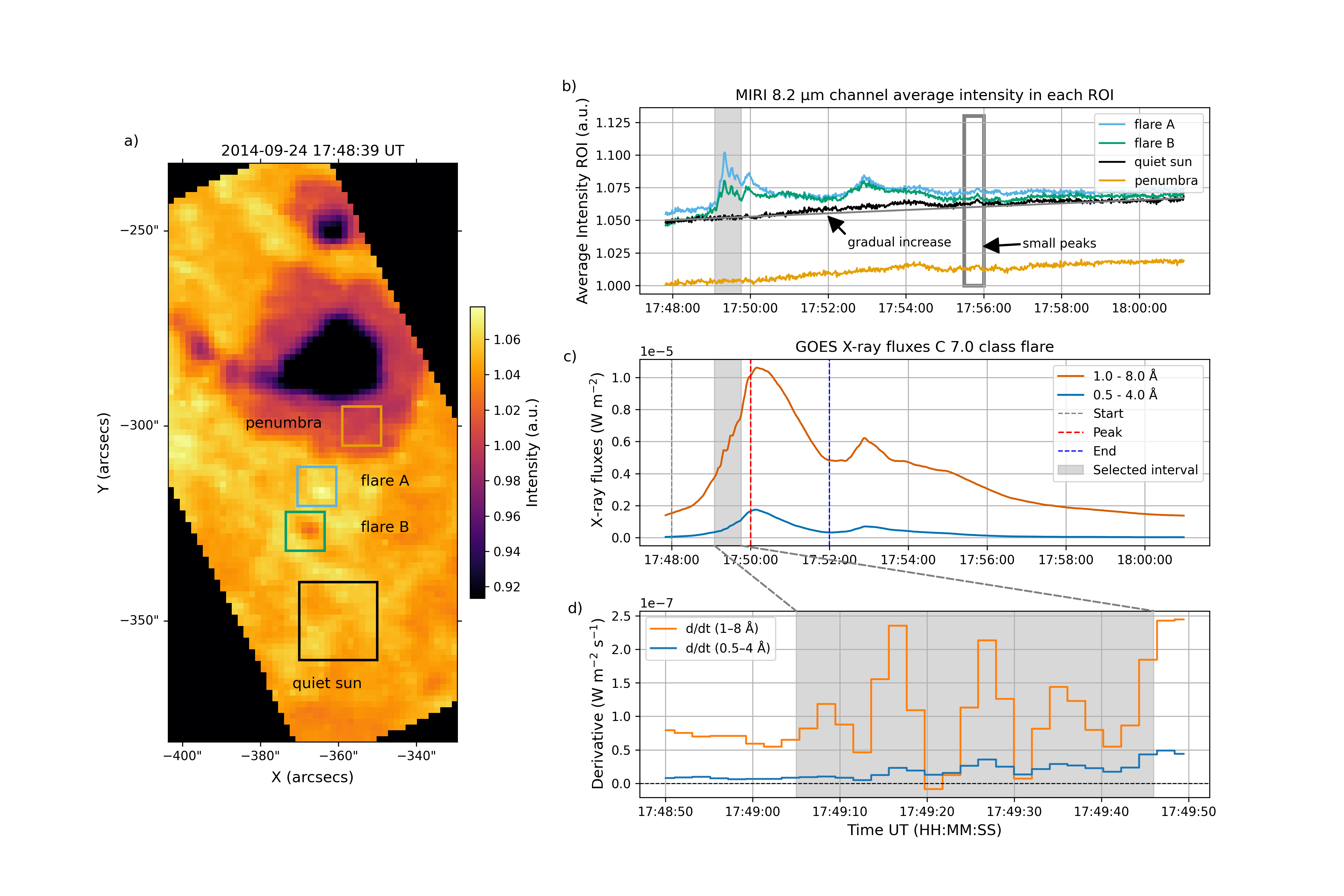}
    \caption{Panel (a): the field-of-view of a sunspot within the active region 12172 captured by the McMath/Pierce East Auxiliary telescope in the \SI{8.2}{\micro\meter} channel. The penumbral region and the quiet sun regions are shown by the yellow and black boxes, respectively. The flare footpoints are shown as blue and green boxed regions, labelled `flare A' and `flare B'. The averaged intensities in these regions are used for further analysis in other panels. Panel (b): Averaged intensity profiles in the 8.2~$\mu$m channel obtained by MIRI in various regions highlighted as coloured boxes in panel a. The grey box indicates small, repeated peaks and a gradual increase over time across all regions, likely due to the effects of Earth's atmosphere. The shaded area indicates the time interval used for calculating the time lags. Panel (c): GOES X-ray fluxes recorded for the C7.0 flare in the 1.0 - 8.0~{\AA} (orange) and 0.5 - 4.0~{\AA} (blue) channels. The three vertical dashed lines indicate the start, peak and end times of the GOES X-ray flare, respectively. Panel (d): the derivative of the SXR flux, which, according to the Newpert effect, is related to the energy depositions. Three peaks can be seen in this interval (17:49:05-17:49:46~UT), just as those observed in the mid-IR emission.}
    \label{fig:1e}
\end{figure*}

We studied the C7.0 class X-ray flare\footnote{\url{https://solarmonitor.org/data/2014/09/24/meta/noaa\_events\_raw\_20140924.txt}} that occurred in Active Region (AR) 12172 (Hale class $\beta\gamma$)\footnote{\url{https://solarmonitor.org/?date=20140924}} on Sept. 24, 2014 (S11 E22). The Solar Object Locator (SOL; \citealt{Leibacher2010}) for this event is SOL2014-09-24T17:50. Fig.~\ref{fig:1e} (panel c) shows the X-ray fluxes for this flare observed by the \textit{Geostationary Operational Environmental Satellite} (GOES-15) in the 1.0 - 8.0~{\AA} and 0.5 - 4.0~{\AA} channels. The flare started at 17:45~UT, peaked at 17:50~UT, and ended at 17:52~UT; these timings are shown as dashed vertical lines.

We studied the mid-IR observations of this flare taken by the McMath/Pierce  \SI{0.81}{\meter} East Auxiliary telescope at the National Solar Observatory\footnote{\url{https://noirlab.edu/public/programs/kitt-peak-national-observatory/mcmath-pierce-solar-telescope/}}, \citep{Pierce1964}, located at the Kitt Peak National Observatory, Arizona, USA. Using the MIRI instrument, the telescope captured this flare in the two mid-IR continuum channels, centred at 5.2 and \SI{8.2}{\micro\meter} with a time resolution of \SI{1}{\second}, a spatial resolution of \SI{0.76}{\arcsecond} per pixel, and a diffraction limit of \SI{2.55}{\arcsecond} at the centroid response wavelength of \SI{8.2}{\micro\meter}, and \SI{1.62}{\arcsecond} at \SI{5.2}{\micro\meter}.

This is an open, all-reflecting, heliostat system with a \SI{0.76}{\meter} aperture and broadband transmission from the ultraviolet to the infrared, as limited by atmospheric conditions \citep{Penn2016SPECTRALMID-INFRARED}. The telescope's f/50 optical design produces a large solar image at prime focus, where a reimaging system using off-axis parabolic mirrors projects it onto a Quantum Well Infrared Photodetector (QWIP).

According to \cite{Penn2016SPECTRALMID-INFRARED} the two-channel QWIP detector employed is based on a GaAs substrate integrated with a ISC0006 silicon readout circuit  \citep{Bundas2006}, with spectral response channels tuned for 4.2–$\SI{6.2}{\micro\meter}$ (called $\SI{5.2}{\micro\meter}$) and 7.0–$\SI{9.3}{\micro\meter}$ (called $\SI{8.2}{\micro\meter}$). They find, after accounting for atmospheric transmission under nominal conditions, that the effective central wavelengths shift to approximately 5.2 and $\SI{8.2}{\micro\meter}$. Each pixel samples the same image region in both channels, with slightly offset exposure times differing by a few milliseconds, which is negligible at the observational cadence of \SI{1}{Hz} used. 

Fig.~\ref{fig:1e} (panel a) shows a field-of-view (FOV) of a sunspot associated with AR~12172 captured in the \SI{8.2}{\micro\meter} channel. The blue and green boxed regions indicate the flare footpoints observed at 17:48:29~UT, and we named them `flare A' and `flare B', respectively. We have chosen two reference regions, the penumbral region of the sunspot, located at the north of `flare A' footpoint region and the quiet sun region, located at the south of the `flare B' footpoint region. We highlight these regions with the yellow and black boxes, respectively. These regions were further used to study the intensity variations.

\begin{figure*}
    \centering
    \includegraphics[trim=0cm 1.0cm 0cm 1.0cm, width=1\textwidth]{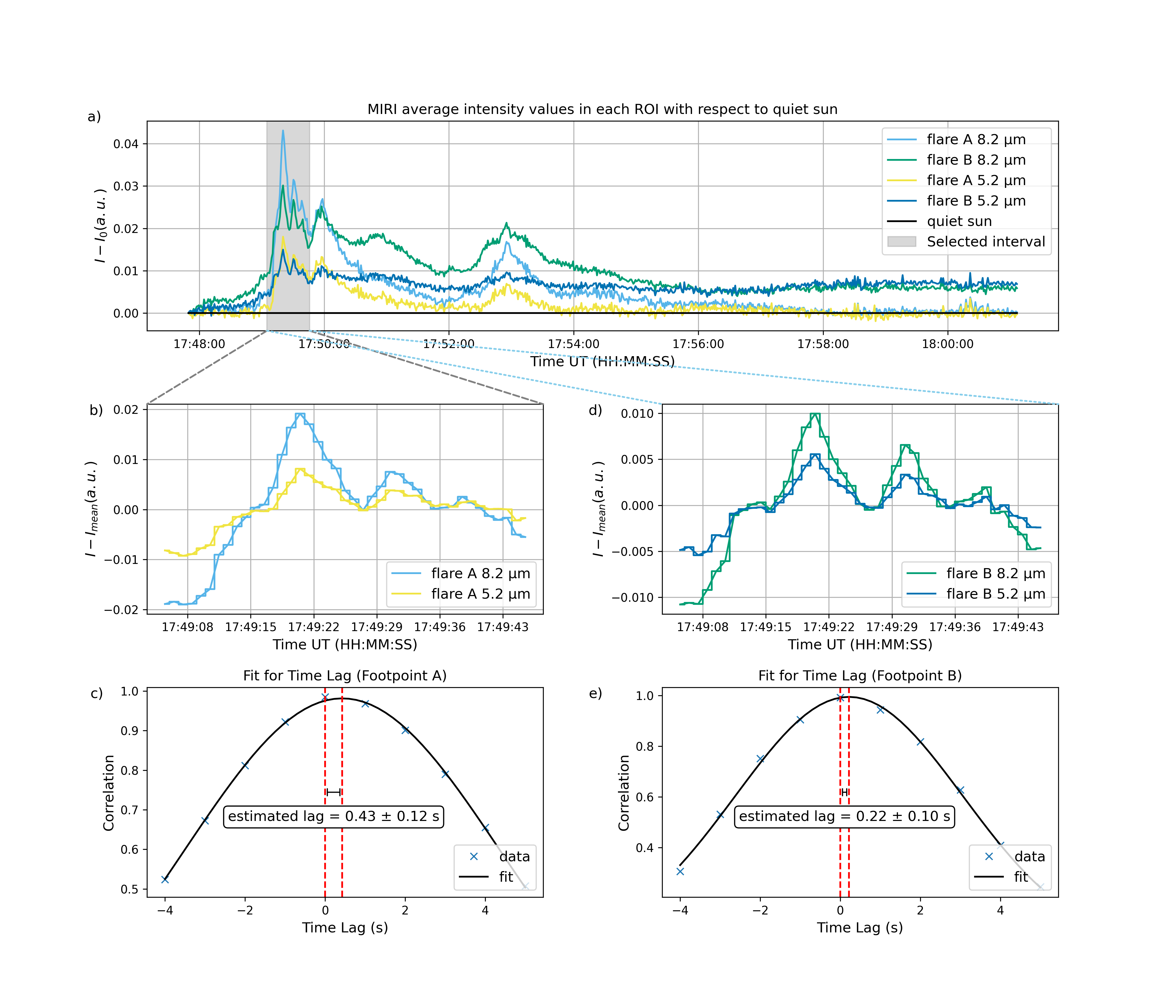}
    \caption{Panel (a): Corrected intensity profiles after subtracting the mean quiet Sun intensity in each channel at the respective flare footpoints. The shaded area indicates the time interval (17:49:05-17:49:46~UT) used to calculate the time lags between the signals in the following panels. Panel (b): The two correlated signals for the footpoint A. Panel (c): Estimated time lag for the footpoint A between the signals from the two channels observed by MIRI. The value obtained from fitting the correlation function is \SI{0.43}{\second} with $\sigma = \SI{0.12}{\second}$. Panel (d) The two correlated signals for the footpoint B. Panel (e): Estimated time lag for the footpoint B between the signals from the two channels observed by MIRI. The value obtained from fitting the correlation function is \SI{0.22}{\second} with $\sigma = \SI{0.10}{\second}$. For the panels c) and e) the vertical red (dotted) lines mark the time lag equal to zero and the maximum value found in the Local CCF in each case. The distance between these two lines represents the lag between the signals.}
    \label{fig:figura2}
\end{figure*}
\subsection{Time evolution of the flare and other regions}
\label{subsec:flare_evolution}

The temporal evolution of flare footpoints, `A and B', quiet sun and the penumbral region in the 5.2 and \SI{8.2}{\micro\meter} channel images was studied thoroughly by taking the averaged intensities for the boxed regions shown in panel a of Fig.~\ref{fig:1e}. The resulting light curves are shown in panel b. 

Additionally, panels c and d show the soft X-ray (SXR) flux measured by GOES-15 and its time derivative. This time derivative is known to correlate with the HXR flux, which in turn is an indicator of non-thermal particles \citep{Neupert1968ComparisonFlares, Steyn2020TheModel}. 
A clear correspondence can be identified between the peaks of the SXR derivative and the emissions observed in both mid-IR channels. Within the analysed interval, three main signals are evident, each associated with distinct injections of non-thermal energy. The normalised HXR time series from the Reuven Ramaty High Energy Solar Spectroscopic Imager (RHESSI), analysed by \cite{Penn2016SPECTRALMID-INFRARED}, shows the same behaviour.

Some of the observed intensity variations could be attributed to interference from the Earth's atmosphere. For example, small intensity peaks that appear consistently across all studied regions suggest a common, non-intrinsic source of variation. Moreover, during the observation period, a gradual increase in intensity is observed across all regions, as shown in Fig.~\ref{fig:1e}. 

Assuming that the average energy flux from the Sun remains constant in the quiet Sun region, we can get the intensity values relative to that mean level. This approach helps mitigate both the spurious peaks and the long-term gradual drift. 
The resulting intensity variations across both channels and their respective footpoints are shown in Fig.~\ref{fig:figura2}; all lag calculations are performed using these curves.

\section{Time-lag analysis}
\label{sec:analysis_results}
\subsection{Lag estimation using the Cross-correlation (CCF) }
\label{subsec:lag_estimation}

Although the cadence of the instrument used is \SI{1}{\Hz}, the lag between the signals can be calculated using sub-resolution techniques such as shown by  \cite{Simoes2024PreciseObservations}. The authors used the standard definition of the CCF to measure the lag from the two spatially resolved infrared sources of the flare SOL2014-09-24T17:50. In this calculation, the peak of the CCF  was fitted with a Gaussian function, allowing the position of the maximum of this function (i.e the lag) to be estimated. 

\begin{figure}
    \centering
    \includegraphics[width=1\linewidth]{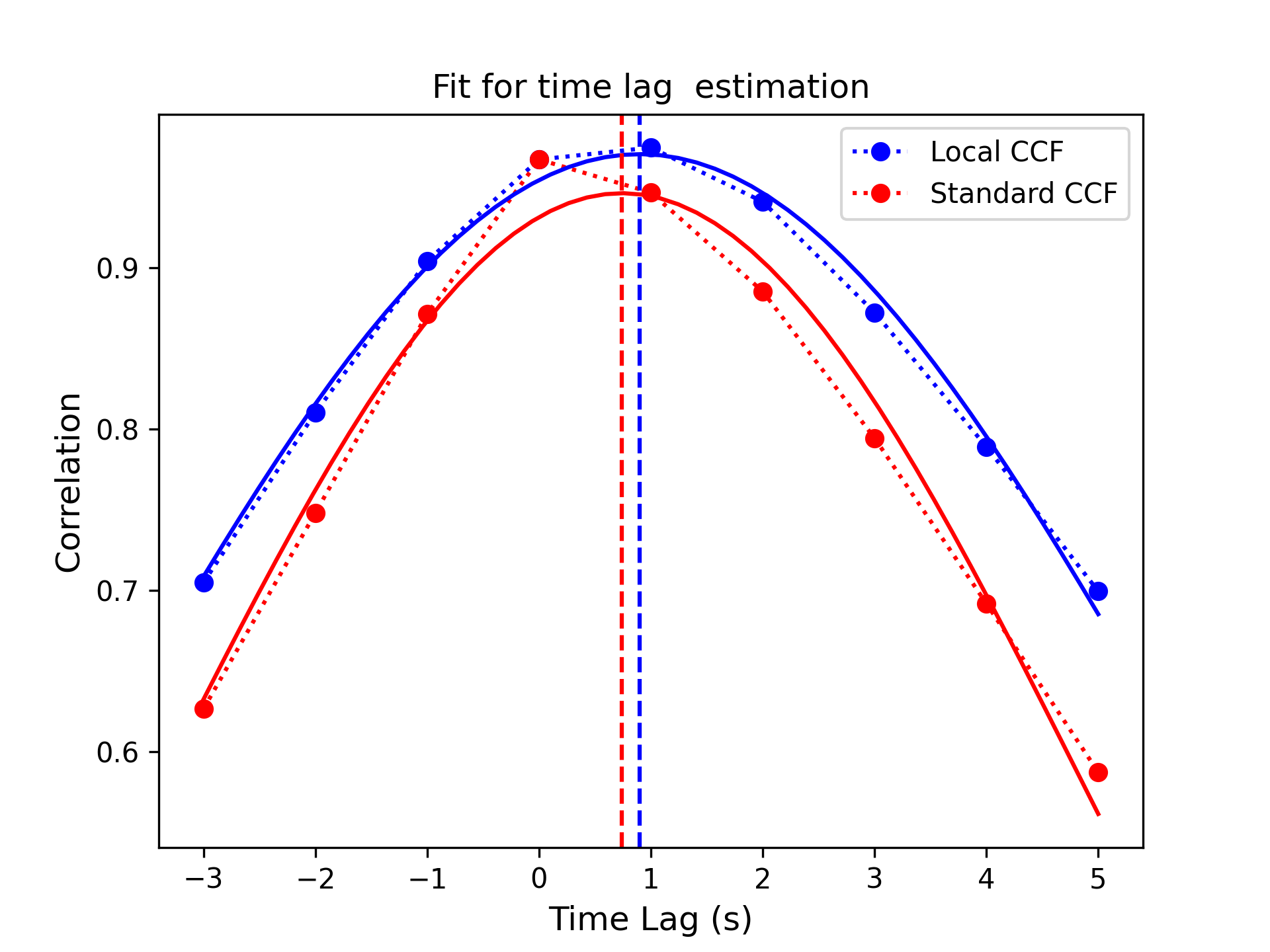}
    \caption{For the flare analysed by Simões et al. (2024), the estimated value for the lag using the local CCF is \SI{0.88}{s} (blue), instead of the reported \SI{0.75}{s} (red).}
    \label{fig:4}
\end{figure}

From the estimation, it is possible to notice that near the maximum, the CCF estimate takes a triangular shape (see the red dotted line) around the maximum in Fig.~\ref{fig:4}, which deviates from the assumed Gaussian form.

Regarding this behaviour, \cite{Welsh1999OnNuclei} argues that the normalisation of the standard definition of the CCF  guarantees that the CCF is always bounded by  $\pm 1$. Still, the normalisation used is only an asymptotically unbiased estimator of the correlation function. Its use introduces a well-known bias toward zero, which grows worse with increasing lag and results in a triangular-shaped reduction of the CCF, underestimating the lag of the peak of the CCF (details of the standard CCF definition can be found in Appendix \ref{A1}). The use of the local correlation function has been proposed as an alternative approach to mitigate lag underestimation, as this definition reduces the bias towards zero \citep{Welsh1999OnNuclei}. The formal definition of this function is also provided in Appendix \ref{A1}.

As an illustration of the difference between the standard and local CCFs, we repeat the lag analysis of \cite{Simoes2024PreciseObservations}. The results are shown in Fig.~\ref{fig:4}. The standard CCF gives \SI{0.75}{s}, confirming their analysis, but  the value calculated using the local CCF is somewhat larger, at \SI{0.88}{s}.

Although \cite{Simoes2024PreciseObservations} analyses the lag from the two spatially resolved infrared sources in each channel separately; in the following section, we here analyse the delays between the two channels (5.2 and 8.2 $\mu$m) for each spatial point.

\subsection{Time delay between 5.2 and 8.2 $\mu$m channels}
\label{subsec:time_delay_analysis}

We have applied the local CCF to determine the lag between the arrival times of signals in each channel (at each footpoint).
The results in Fig.~\ref{fig:figura2}, panels c) and e) show that there is a delay between the channels of \SI{0.43 +- 0.12}{\second} for the northernmost footpoint, and a delay of \SI{0.22 +- 0.10}{\second} for the southernmost footpoint. 
In both cases, the signal from the \SI{8.2}{\micro\meter} channel peaks first compared to the \SI{5.2}{\micro\meter} channel. 

We estimated the uncertainties associated with noise effects in the measured time lags using a Monte Carlo analysis. To do this, we selected a post-flare noise-dominated interval 17:59:00 and 18:00:00 UT and computed the standard deviation of the noise in each signal. Using this value, we generated Gaussian noise (both positive and negative) and randomly added it to the original time series. The resulting time series were then processed with the same correlation method, yielding a time lag for each realization. This procedure was repeated 1000 times, and the standard deviation of the resulting time lags was adopted as the reported uncertainty of the measurement.

\cite{Simoes2017FormationFlares} showed that at different IR wavelengths the continuum opacity contribution of the two main sources, H and H$^{-}$ (negative hydrogen ion) opacity \citep{Heinzel2012Optical-to-radioFlares}, varies somewhat. So the observed delay between the two channels may be due to the different evolution with time of these opacities, as a flare proceeds.
We will next investigate the delay by calculating the emission in each of these channels using the method suggested by \cite{Simoes2017FormationFlares}, which uses the results of the 1D radiation hydrodynamics code RADYN, to determine the conditions of the solar atmosphere after the energy injection of the flare and then based on these parameters calculate the radiative transfer. 

\section{Calculating infrared emission}
\label{sec:IR_radiation_flare}

To model infrared emission during a solar flare, we consider the radiative transfer equation, whose general solution is given by:

\begin{equation}
I_\nu(s_0) = \int_{0}^{s_0} j_\nu(s) \exp\left[-\int_s^{s_0} k_\nu(s') \, \mathrm{d}s'\right] \mathrm{d}s,
\end{equation}

where $I_\nu(s_0)$ is the specific intensity of the radiation emitted at the frequency $\nu$ observed at position $s_0$, $j_\nu$ is the emission coefficient, and $k_\nu$ is the absorption coefficient. This expression accounts for the contribution of each layer (at a height $s$) along the line of sight, integrating from deeper atmospheric layers ($s=0$) towards the observer ($s=s_0$). In our case, for the F-CHROMA grid results \citep{Carlsson2023TheModels}, zero is taken as \SI{90}{\kilo\meter} below the height of $\tau_{500 \mathrm{~nm}}=1$. 

Considering the definition of the optical depth
\begin{equation}
\tau_\nu(s)=\int_s^{s_0}  k_v\left(s^{\prime}\right) \mathrm{d} s^{\prime}   ,
\end{equation}
and the definition of the contribution function (CF) as: 
\begin{equation}
CF(s)= j_{\nu} e^{- \tau_\nu(s) },
\label{eq:03}    
\end{equation}
then:
\begin{equation}
I_\nu\left(s_0\right)=\int_{0}^{s_0} CF(s) \mathrm{d} s.
\end{equation}

This equation states that the observed intensity is composed of the energy carried by photons emitted from all locations in front of $s_0$ that have not been absorbed or scattered into other directions on the way to the observer. 

The emission coefficient can be written on terms of the source function ($S_\nu$) as $j_{\nu} = k_\nu S_\nu$ and, in the formation of the infrared continuum, we can assume that this function is given by the local Planck function ($S_v=B_v(T)$), with $T$ being the electron temperature.

\begin{equation}
  B_\nu(T)=\frac{2 h \nu^3}{c^2} \cdot \frac{1}{e^{\frac{h \nu}{k_{\mathrm{B}} T}}-1} . 
\end{equation}

According to \cite{Heinzel2012Optical-to-radioFlares} and references there, in the chromosphere, the main source of opacity for the infrared continuum is ion-free-free continuum, and in the temperature minimum region and below, the opacity is dominated by neutral free-free opacity.

Following the process described by \cite{Simoes2017FormationFlares}, the hydrogen free-free $k_\nu(\mathrm{H})$ absorption coefficient (in \si{cm^{-1}}) can be written as:
\begin{equation}
k_\nu(\mathrm{H})=3.7 \times 10^8 T^{-1 / 2} n_{\mathrm{e}} n_p \nu^{-3} g_{\mathrm{ff}},   
\label{eq:01}
\end{equation}
where $n_{\mathrm{e}}$ and $n_p$ are the electron and proton densities respectively,  $T$ is the kinetic (or electron) temperature,  and $g_{\mathrm{ff}}$ is the Gaunt factor,  which is calculated using a bilinear interpolation of the factors calculated in Table 3 of \cite{vanHoof2014AccurateFactors}. The $n_e$ includes contributions from hydrogen, helium, and metals.

In the lower atmosphere, around the temperature minimum region and below, H$^{-}$ 
free-free opacity dominates, and this can be calculated (in \si{cm^{-1}}) as:
\begin{equation}
k_\nu\left(\mathrm{H}^{-}\right)=\frac{n_{\mathrm{e}} n_{\mathrm{H}}}{\nu}\left(A_1+\left(A_2-A_3 / T\right) / \nu\right),
\label{eq:02}
\end{equation}

where $n_{\mathrm{H}}$ is the neutral hydrogen density, and the numerical coefficients are $A_1=1.3727 \times 10^{-25}, A_2=4.3748 \times 10^{-10}$, and $A_3=2.5993 \times 10^{-7}$. 

Finally, the total absorption coefficient is: 
\begin{equation}
k_\nu=\left[\kappa_v(\mathrm{H})+\kappa_v\left(\mathrm{H}^{-}\right)\right]\left(1-e^{-h v / k_{\mathrm{b}} T}\right)    
\end{equation}
with the term $\left(1-e^{-h \nu / k_b T}\right)$ being the correction for stimulated emission, where $h$ and $k_{\mathrm{b}}$ are the Planck and Boltzmann constants.

\section{Modeling the delay using RADYN simulations}
\label{sec:RADYN_analysis}

The 1D radiation hydrodynamics code RADYN simulates the response of the solar atmosphere to a beam of non-thermal electrons injected at the apex of a coronal loop, assuming time-independent beam properties. The F-CHROMA grid results \citep{Carlsson2023TheModels} include precomputed RADYN simulations\footnote{Available at \href{https://star.pst.qub.ac.uk/wiki/public/solarmodels/start.html}{https://star.pst.qub.ac.uk/wiki/public/solarmodels/start.html}} of the evolution of solar atmospheric conditions, accounting for different parameters of the electron beam generated during flares: the integrated energy flux during the entire energy input period (known as fluence $\mathscr{F}$), the spectral index ($\delta$), and the low-energy cut-off ($E_{\mathrm{c}}$) of the electron beam. In all models, the authors employed a triangular shape for the beam energy flux as a function of time, with a linear increase in beam flux from zero at $t=\SI{0}{\second}$ to a maximum at $t=\SI{10}{\second}$ and then a linear decrease to zero beam flux at $t=\SI{20}{\second}$. 

The population of each species, needed to calculate the opacity coefficients according to Equations \ref{eq:01} and \ref{eq:02} at each time and height, was derived from the RADYN outputs. This model grid enables us to explore different heating scenarios and assess their influence on the time delay between mid-IR emission peaks observed at different wavelengths. The range of parameters used in this paper is shown in Table \ref{tab:radyn}.  

\begin{table}
\caption{Selected parameter values used in this paper from the RADYN grid by  Carlsson et al. (2023).}
\begin{tabular}{lc}
\hline \hline Parameter & Values \\
\hline$\delta$ & $3,4,5$ \\
$\mathscr{F}$ & $1 \times 10^{11}, 3 \times 10^{11}$, \SI{1e12}{erg.cm^{-2}} \\
$E_{\mathrm{c}}$ & $10,15,20$ $\mathrm{keV}$ \\
\hline
\end{tabular}
\label{tab:radyn}
\end{table}

\subsection{Time lag from observationally-derived beam parameters}
The first scenario considered in this study aims to simulate the conditions of an electron beam similar to that inferred for the flare analysed in the previous sections.
\citet{Simoes2024PreciseObservations} conducted a detailed analysis of RHESSI data \citep{Lin2002TheRHESSI} to determine the properties of accelerated electrons at the flare footpoints, labelled A and B in this paper. Using these observations during the impulsive phase of the event (17:49:05--17:49:33~UT), they 
derived spectral indices $\delta_A = 2.83 \pm 0.13$ and $\delta_B = 3.11 \pm 0.15$, for the northern and southern footpoints, respectively. Based on spectral fitting, the low-energy cut-off of the electron distribution was estimated to be $15 \pm 30$~keV. 
They also estimated the energy fluence delivered by non-thermal electrons to each footpoint, 
deriving energy fluxes in the ranges $0.5 \times 10^{10} < F_A < 1.3 \times 10^{10}~\mathrm{erg\,s^{-1}\,cm^{-2}}$ and $0.9 \times 10^{10} < F_B < 1.9 \times 10^{10}~\mathrm{erg\,s^{-1}\,cm^{-2}}$. These values are considered lower limits due to the limited spatial resolution of the infrared observations. 
Considering these results, and bearing in mind that the RADYN F-CHROMA grid assumes a 20\,s triangular injection the closest available simulation has $\delta = 3$, $E_{\text{c}} = \SI{15}{keV}$ and $\mathscr{F} = \SI{1e11}{\erg\per\square\centi\meter}$.
The longer energy injection timescales in the F-CHROMA RADYN simulations models may enhance hydrodynamic responses compared with the shorter injections used by \cite{Simoes2024PreciseObservations}, although the same qualitative wavelength-dependent timing is also found in simulations with short injections times.

\begin{figure*}
    \centering
    \includegraphics[width=1\linewidth]{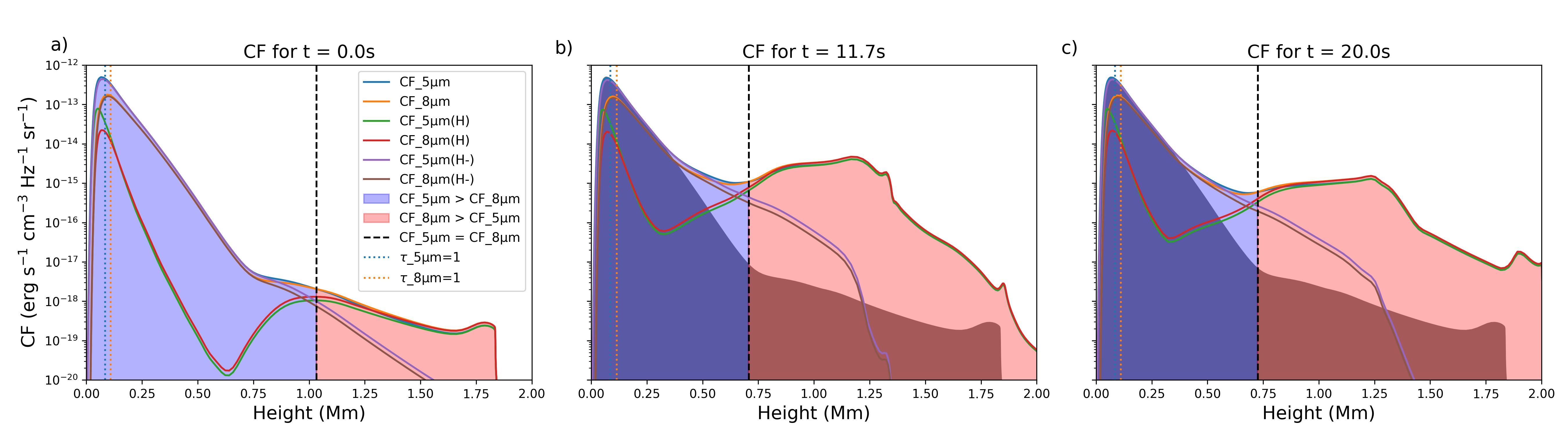}
    \caption{Variation of the contribution function at three instants during the flare. Panel (a): the initial conditions; Panel (b): the conditions at the time of maximum emission in the \SI{8}{\micro\metre} channel, since the variation in intensity is with respect to the initial values. The initial CF is shown in shaded form. Panel (c): the conditions at the time when the RADYN simulation stops injecting energy. The animation associated with this figure is available online.}
    \label{fig:CFs}
\end{figure*}

The Fig.~\ref{fig:CFs} shows the contribution function (equation \ref{eq:03}) for different time points, calculated using the RADYN output parameters as a function of height. (An animation of how these vary over time is available online). It is possible to distinguish that the contribution function is greater in the \SI{5.2}{\micro\meter} band in the deeper regions of the atmosphere, whereas it is smaller than the \SI{8.2}{\micro\meter} contribution function in the higher layers.

It is also observed that the flare emission detected in both cases forms in an optically thin atmosphere ($\tau \ll 1$), in agreement with \cite{Simoes2017FormationFlares}. During the flare, the region where the \SI{8.2}{\micro\meter} CF dominates advances towards deeper layers of the atmosphere, reaching the minimum depth of around $\SI{0.74}{\mega\meter}$ at $\SI{6.6}{\second}$. Afterwards, although the flare energy deposition in the simulation continues until $t = \SI{20}{\second}$, this region returns to the higher layers of the atmosphere.

\begin{figure*}
    \centering
    \includegraphics[width=1\linewidth]{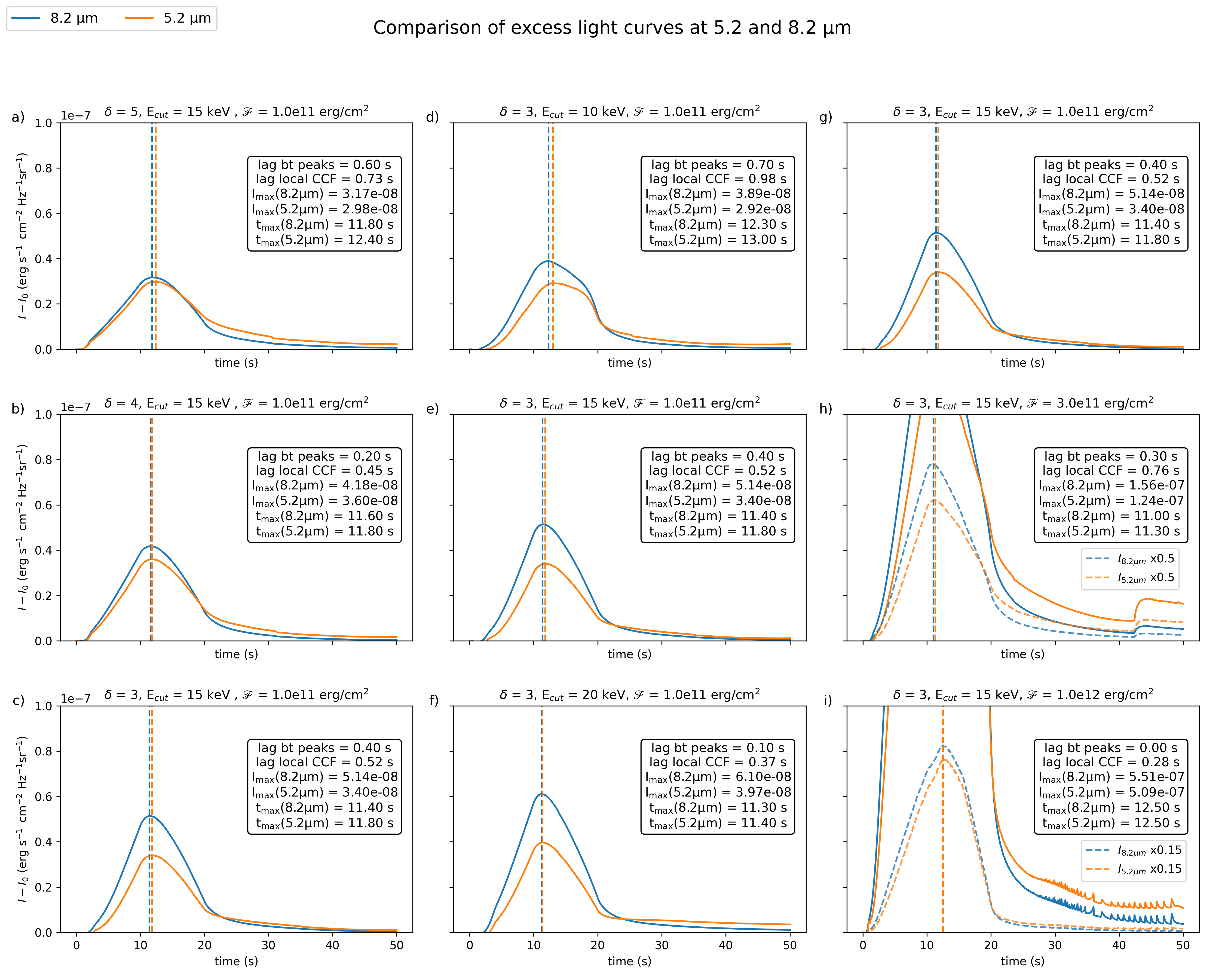}
    \caption{Intensity variations in both channels for different scenarios. The first column contains cases in which the spectral index varies; the second column contains cases in which the cut-off energy values vary; and the third column contains cases in which the energy fluence values vary. The lag (in seconds) between the maximum emission peaks taken directly from the RADYN model results is shown in the legend, as well as the lag value using the local CCF between the signals taking a sampling with a cadence of 1s (similar to the observations described in section \ref{subsec:time_delay_analysis}). Finally, it also includes the maximum intensities at the emission peaks (in $\text{erg } \mathrm{s}^{-1} \mathrm{~cm}^{-2} \mathrm{~Hz}^{-1} \mathrm{sr}^{-1}$). For panels h) and i) a scaled version in dotted lines has been added.}
    \label{fig:radyn_outs}
\end{figure*}

By integrating the contribution function, the variation of the intensity in each channel was calculated (Fig.~\ref{fig:radyn_outs} panel c). For the case of \SI{8.2}{\micro\meter} the emission peak occurs at $t=\SI{11.4}{\second}$ while for \SI{5.2}{\micro\meter} it occurs at $t= \SI{11.8}{\second}$. In this scenario, the time lag between the peaks is approximately \SI{0.4}{\second}. Using the local CCF for this signal and assuming a sampled version with a cadence of \SI{1}{\second}, the estimated lag is \SI{0.52}{\second}, which is consistent with the value found experimentally in the previous section.

\subsection{Dependence of time lag on beam parameters}
To determine the effects of time lag resulting different electron beam parameters, different scenarios were repeated varying one parameter while holding the others constant. The results can be seen in Fig.~\ref{fig:radyn_outs}.

In all cases, both channels show a very similar temporal evolution, with the emission at \SI{5.2}{\micro\meter} consistently peaking later than at \SI{8.2}{\micro\meter}. The measured lag varies between 0 and \SI{0.7}{\second}, depending on the characteristics of the electron beam. Panels (a–c) indicate that harder spectra (lower $\delta$) tend to produce larger delays between the two channels. Panels (d-f) shows that increasing the low-energy cut-off from 10 to 20 keV results in a reduction of the lag (from \SI{0.7}{\second} to \SI{0.1}{\second}). Variations in the fluence (panels g–i) also affect the time lag and the relative amplitudes of the signals: higher fluence values enhance the intensity of both channels and shorten the delay between their peaks. These results suggest that the lag between mid-IR emissions is sensitive to both the spectral index and the energy deposition parameters of the electron beam.

\begin{figure}
    \centering
    \includegraphics[width=1\linewidth]{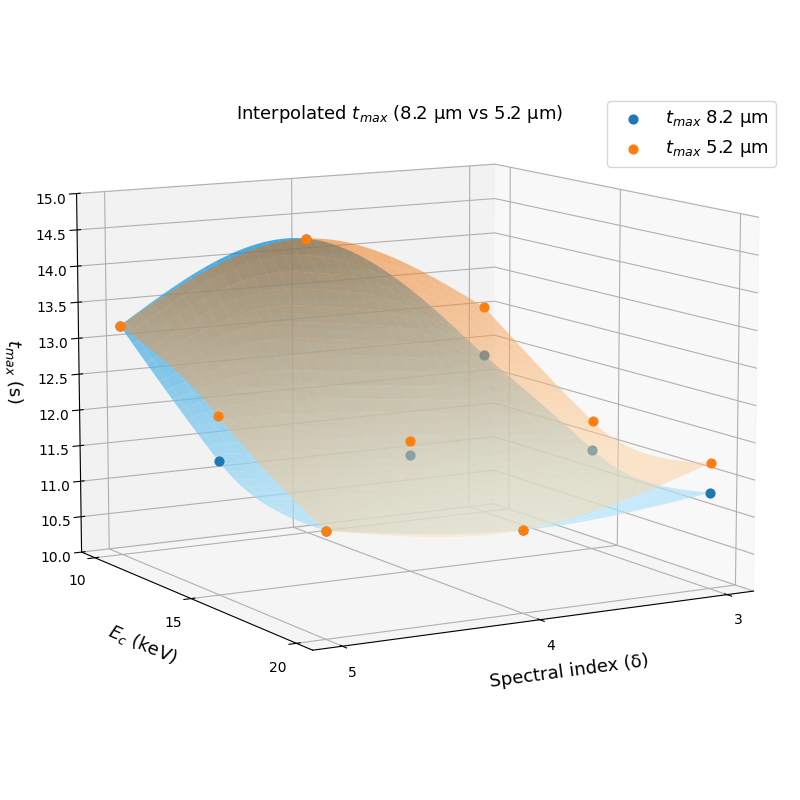}
    \caption{Interpolated values of $t_{\mathrm{max}}(\lambda)$ for different spectral indices and cut-off energies, for a fixed fluence of $\mathscr{F} = \SI{1e11}{\erg.\centi\meter^{-2}}$. Each surface corresponds to a wavelength channel, illustrating how the time required to reach the emission peak varies with the beam parameters; the z-axis indicates the time required to reach maximum emission from the moment the energy injection begins.}
    \label{fig:tmax_inter}
\end{figure}

For an electron beam with a given temporal profile --- in our case, a triangular function peaking at $t = \SI{10}{\second}$ and a full width of \SI{20}{\second} --- considering $t=\SI{0}{\second}$ as the moment when the energy input begins, the time at which the emission reaches its maximum in a particular wavelength channel $ t_{\mathrm{max}}(\lambda)$, depends on the fluence, spectral index, and cut-off energy. Fig.~\ref{fig:tmax_inter} illustrates the behaviour of this dependency at a fixed fluence. In this case, we interpolate between the $t_{\mathrm{max}}$ values obtained from the outputs of the RADYN simulations. It can be seen that $t_{\mathrm{max}}$ decreases for higher cut-off energies. In this representation, the distance along the z-axis corresponds to the time necessary to reach the maximum peak emission.

\begin{figure}
    \centering
    \includegraphics[width=1\linewidth]{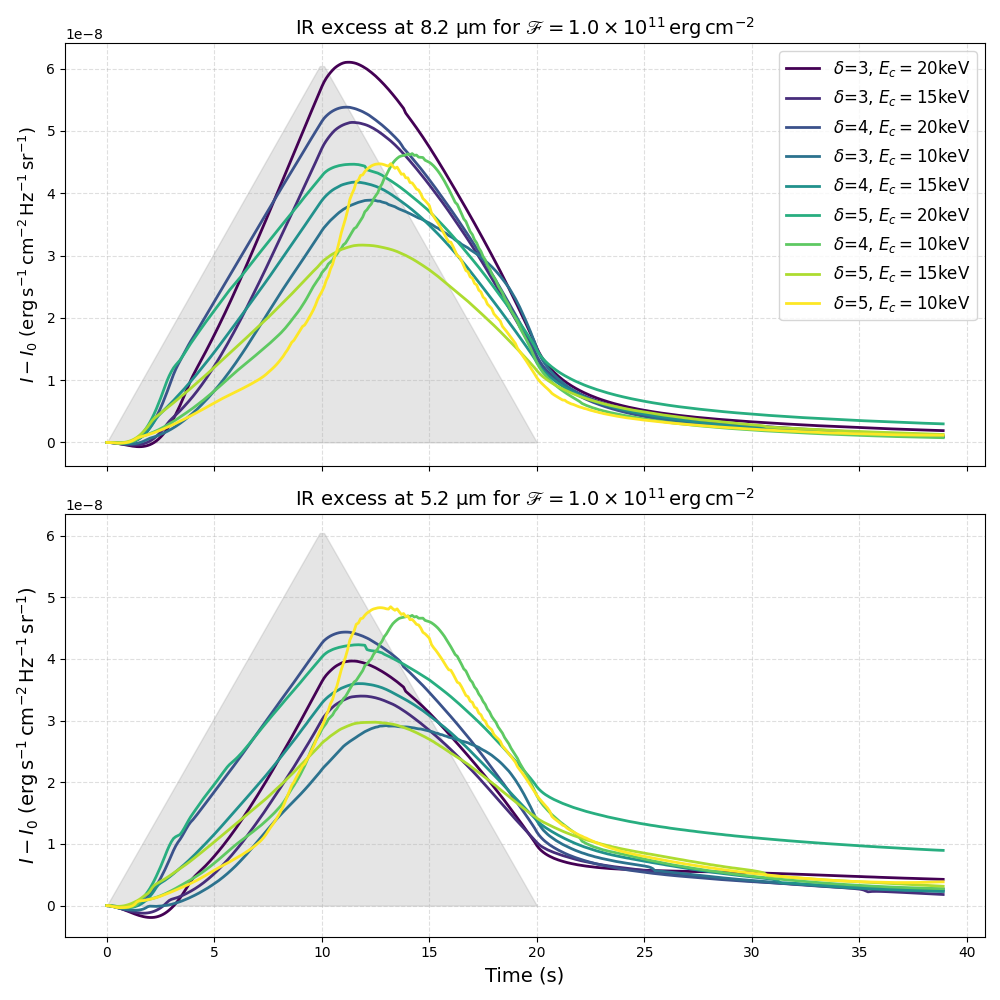}
    \caption{Excess brightness curves for each mid-IR channel in the case of $\mathscr{F} = \SI{1e11}{\erg.\centi\meter^{-2}}$. The colour scale represents the number of electrons at \SI{50}{\kilo\electronvolt} (purple for the largest, yellow for the smallest). The shaded area denotes the temporal profile of energy input. The \SI{5.2}{\micro\meter} emission generally shows a longer delay, while for larger populations of energetic electrons the \SI{8.2}{\micro\meter} emission more closely follows the input profile.}
    \label{fig:exess_curves}
\end{figure}

For a given fluence, the cut-off energy and spectral index are particularly relevant because they are directly related to the penetration depth of the electrons into the solar atmosphere. Fig.~\ref{fig:exess_curves} shows the excess brightness curves for each channel, considering the case of $\mathscr{F} = \SI{1e11}{\erg.\centi\meter^{-2}}$. The colour-coding corresponds to the number of electrons at \SI{50}{\kilo\electronvolt} \citep[following][]{Simoes2017FormationFlares}, with purple indicating the largest and yellow the smallest. This number is significant because it provides an estimate of the amount of high-energy electrons capable of reaching deeper layers of the chromosphere.

In general, the response in the \SI{5.2}{\micro\meter} channel occurs with a larger delay relative to the energy input profile (shown as shaded region), compared to the the \SI{8.2}{\micro\meter} channel. When a large number of energetic electrons are present --- particularly for the \SI{8.2}{\micro\meter} channel --- the mid-IR emission follows the energy input profile more closely.

Depending on the characteristics of the electron beam, the system's response to the energy input can be rapid or delayed. In some cases, the emission peak is reached near the maximum injected power (which occurs at $t = \SI{10}{\second}$), but in some cases, the emission peak occurs practically at the end of the energy input period (at $ t= \SI{20}{\second}$). Analysing the results according to the timing of ($t_{\mathrm{max}}$) at \SI{5.2}{\micro\meter}, it is observed that, in most cases with a delayed response, the \SI{5.2}{\micro\meter} peak exceeds that of \SI{8.2}{\micro\meter}. 

When considering scenarios with the same energy and the same low-energy cut-off, the cases with a higher spectral index ($\delta$) tend to show greater emission at \SI{5.2}{\micro\meter} compared to \SI{8.2}{\micro\meter}, with this difference being more pronounced as ($\delta$) increases. In contrast, cases with low spectral indices show greater emission at \SI{8.2}{\micro\meter}, and the difference between the two channels increases as ($\delta$) decreases.

Scenarios with a higher value of the low-energy  cut-off tend to generate rapid responses, reducing the delay time between the start of the energy input and the maximum emission. Consequently, as the energy cut-off increases, ($t_{\mathrm{max}}$) decreases, indicating that the rapid response occurs sooner after the energy input. In contrast, late responses predominate in conditions with low energy cut-off values.

\begin{table}
\centering
\caption{ Time in seconds for each channel to reach its peak ($t_{\mathrm{max}}$) as a function of spectral index and energy cut-off, for a fluence of $\mathscr{F} = \SI{1e11}{erg.cm^{-2}}$.} 
\label{tab:tmax}
\begin{tabular}{|c|S|S|S||S|S|S|}
\hline
\multirow{3}{*}{\begin{tabular}{c}Energy cut-off\\ {[keV]}\end{tabular}}
& \multicolumn{3}{c||}{\SI{8.2}{\micro\meter}}
& \multicolumn{3}{c| }{\SI{5.2}{\micro\meter}} \\ \cline{2-7}
& \multicolumn{3}{c||}{Spectral index $\delta$}
& \multicolumn{3}{c|}{Spectral index $\delta$} \\ \cline{2-7}
& {3} & {4} & {5} & {3} & {4} & {5} \\
\hline
10 & \cellcolor{celesteoscuro} 12.3 & \cellcolor{celesteoscuro}14.2 & \cellcolor{celesteoscuro}13.2
   & \cellcolor{anaranjado}13.0& \cellcolor{anaranjado}14.2& \cellcolor{anaranjado}13.2\\
\hline
15 & \cellcolor{celesteoscuro}11.4& \cellcolor{celesteoscuro}11.6& \cellcolor{celesteoscuro}11.8
   & \cellcolor{anaranjado}11.8& \cellcolor{anaranjado}11.8& \cellcolor{anaranjado}12.4\\ \hline
20 & \cellcolor{celesteoscuro}11.3& \cellcolor{celesteoscuro}11.1& \cellcolor{celesteoscuro}11.4
   & \cellcolor{anaranjado}11.4& \cellcolor{anaranjado}11.1 & \cellcolor{anaranjado}11.7\\
\hline
\end{tabular}
\end{table}

\begin{table}
\centering
\caption{Time lag results in seconds for different values of spectral index and energy cut-off for a fluence of $\mathscr{F} = \SI{1e11}{erg.cm^{-2}}$.}
\label{tab:combined}
\begin{tabular}{|c|c|c|c|}
\hline Energy cut-off & \multicolumn{3}{|c|}{ Spectral index $\delta$} \\
\cline { 2 - 4 }$[\mathrm{keV}]$ & 3 & 4 & 5 \\
\hline 10 & 0.7 & 0.0 & 0.0 \\
\hline 15 & 0.4 & 0.2 & 0.6 \\
\hline 20 & 0.1 & 0.0 & 0.3 \\
\hline
\end{tabular}
\end{table}

\subsection{Possible implication of the observed IR time lags}
\label{subsec:time_lag_implication}
From the RADYN simulations, the physical origin of the time lags between the two channels can be understood in terms of the ionisation response at different atmospheric depths and the wavelength dependence of the dominant emissivity sources. 

As the IR spectrum is computed in LTE, the source function is closely tied to the local Planck function. However, the mid-IR emissivity and opacity depend only weakly on temperature. During the flare, non-thermal ionisation and subsequent recombination rapidly modify the electron density in the upper chromosphere. Because these layers are relatively tenuous, the fractional change in $n_e$ can be very large, producing a rapid increase in the hydrogen free--free emissivity $j_\nu(H)$. The \SI{8.2}{\micro\meter} channel, which is more sensitive to emission from ionised hydrogen, therefore responds earlier to these changes in electron and proton density. 

In contrast, deeper and denser layers, where H$^-$ emissivity $j_\nu(\mathrm{H}^-)$ dominates, experience smaller relative changes in electron density and evolve more gradually (due to their larger heat capacity and higher density). As a result, the \SI{5.2}{\micro\meter} emission, which receives a larger contribution from these layers, peaks later in time.

 Near the photosphere, $\mathrm{CF}_{8\,\mu\mathrm{m}} < \mathrm{CF}_{5\,\mu\mathrm{m}}$, whereas at greater heights $\mathrm{CF}_{8\,\mu\mathrm{m}} > \mathrm{CF}_{5\,\mu\mathrm{m}}$. During a flare, enhanced ionisation increases the proton and electron densities at progressively greater depths, strengthening $j_\nu(H)$ and pushing the point where $j_\nu(H) = j_\nu(\mathrm{H}^-)$ downward. This behaviour is evident in the contribution function (see Fig. \ref{fig:CFs}).

Because longer wavelengths are more strongly influenced by hydrogen free--free emission formed in rapidly responding upper layers, they react on shorter timescales as the electron density increases rapidly due to non-thermal ionisation. Shorter wavelengths, which retain a relatively stronger contribution from the more gradually changing deeper layers where H$^-$ dominates, exhibit a slower intensity rise.

The properties of the non-thermal electron beam also influence the temporal response of the infrared emission. Harder beams (with more high-energy electrons above \SI{50}{\kilo\electronvolt}) deposit energy more efficiently in the chromosphere and produce a faster increase in electron density through a combination of thermal and non-thermal ionisation of hydrogen. As a result, the infrared emission responds more rapidly when the beam spectrum is harder, as illustrated in Fig.~\ref{fig:exess_curves}, where the emission peaks occur significantly earlier for harder beams.

\section{Conclusions}
\label{conclusion}

Mid-IR observations obtained during the flare revealed, for the first time, a measurable time lag between two mid-IR emission channels (\SI{8.2}{\micro\meter} and \SI{5.2}{\micro\meter}) at each footpoint. We found a \SI{0.45}{\second} lag at footpoint A and a \SI{0.28}{\second} lag at footpoint B. In both cases, the \SI{8.2}{\micro\meter} emission peaks first. This lag was measured using the local cross-correlation function. This approach offers advantages over the standard cross-correlation function, as it reduces the bias toward zero that is commonly present in global methods. Using this approach, we also confirm the lag from the two spatially resolved infrared sources reported by \cite{Simoes2024PreciseObservations}, only in this case using the local CCF we obtained a slightly larger value of \SI{0.88}{\second}.

We investigated possible physical causes of this delay by applying the method proposed by \citet{Simoes2017FormationFlares} to estimate infrared emission from the one-dimensional radiation-hydrodynamics code RADYN and by considering the different scenarios of the F-CHROMA grid \citep{Carlsson2023TheModels}.

According to these calculations, the observed lag between peaks is related to the fact that, for a given energy input, the infrared emission at longer wavelengths responds more rapidly than at shorter wavelengths (this explains why the peak at \SI{8.2}{\micro\meter}  occurs earlier than that at \SI{5.2}{\micro\meter} in the considered models). This faster response arises because, in the outer chromospheric layers, the contribution function is dominated by the ionised H density, which changes rapidly during energy injection and in these layers, the contribution from \SI{8.2}{\micro\meter} is greater than that from \SI{5.2}{\micro\meter}. On the other hand, the H$^{-}$ density becomes more relevant in the lower layers, where, on the contrary, the contribution function for the \SI{5.2}{\micro\meter} is greater. In general, shorter wavelengths, originate in deeper layers, which exhibit a delayed response relative to the energy deposition.

In the simulations, for each wavelength channel, it is possible to determine the time of maximum emission as a function of the spectral index, the cut-off energy, and the total energy fluence. This allows us to identify which scenarios correspond to a faster response that closely follows the energy input, and which ones exhibit a larger delay. Subsequently, it becomes possible to make theoretical comparisons between different channels and determine the time lag between them.

For a given energy fluence, there is also a correlation between the ratio of emission peak intensities and the number of high-energy electrons. Specifically, when the number of energetic electrons is small, the peak intensities at both wavelengths are similar. This occurs because a higher number of energetic electrons allows the energy to reach deeper chromospheric layers, where, as shown in Fig.~\ref{fig:CFs}, the contribution function is dominated by emission at \SI{5.2}{\micro\meter}.

Since, during a flare, the response time of a particular channel depends on the wavelength, this time-delay analysis technique provides a new diagnostic tool to probe the consequences of energy deposition processes across different chromospheric layers. As discussed above, different infrared channels contribute more significantly in specific chromospheric regions — for example, in the upper layers, the contribution from the \SI{8.2}{\micro\meter} channel is larger than that from \SI{5.2}{\micro\meter}, while the latter is comparatively more sensitive to emission from deeper layers.

Our modeling assumption that the energy is transported through the solar chromosphere by an electron beam appears at odds with \citet{Simoes2024PreciseObservations}'s finding of a time lag between the two footpoints that is too long to be explained by an electron beam carrying the flare energy through the corona, unless some rather extreme assumptions are made. Note  that \citet{Simoes2024PreciseObservations}  prompts questions about how the energy \emph{arrives at} the chromosphere, whether by particle beams, magnetic waves, thermal conduction, or a mix of the three. However, it is undeniable that fast electrons are \emph{present in} the chromosphere in this event, demonstrated by the HXR sources co-spatial with the IR sources. And the deduced electron energy flux means they will have a major role in transporting energy through the chromosphere. So without assuming anything about energy transport through the corona, we can nonetheless take the electron properties derived from HXR analysis of chromospheric sources as the starting point to motivate our choice of RADYN simulations of chromospheric electrons.  This is not completely consistent because: (i) the RADYN simulations start the electrons at the top of a coronal loop, not at the top of the chromosphere and (ii) the RADYN simulations used assume electrons with a quite strongly-beamed distribution, whereas this might not be the case if electrons were accelerated somewhere in or much nearer the collisional chromosphere. However, since the evolution of the electron beam will be dominated by what happens in the much denser chromosphere we expect that the effect of ignoring the corona will be minimal. We would hope in the future to repeat the model-data comparison with models for other modes of energy transport \citep[e.g.][]{2025ApJ...986...73L}.

The analysis here has demonstrated some of the capabilities of the mid-IR. These observations can have excellent contrast for flare emissions, with intrinsically high time resolution and the possibility of multiple spectroscopic channels. Higher temporal resolution measurements across multiple mid-IR channels would enable a more detailed understanding of how energy is deposited during a flare.

\section*{Acknowledgements}
M. Rojas-Quesada, acknowledges the support of a PhD Scholarship from the College of Science and Engineering at the University of Glasgow, 
and the Costa Rica Institute of Technology for granting the necessary permissions to undertake PhD studies. L. Fletcher and S. Mulay acknowledge support from UK Research and Innovation’s Science and Technology Facilities Council under grant award number  ST/X000990/1. P.~J.~A.~Sim\~oes acknowledges support from Conselho Nacional de Desenvolvimento Científico e Tecnológico (CNPq) (contract 305808/2022-2) and Fundação de Amparo à Pesquisa do Estado de São Paulo (FAPESP) contract 2022/15700-7.\\

We acknowledge Matt Penn and the GSFC team who developed the instrument and conducted the observations we used. We also would like to acknowledge the F-CHROMA team for providing access to the RADYN F-CHROMA simulation grid, which was essential for this work. The research leading to the development of these models received funding from the European Community’s Seventh Framework Programme (FP7/2007–2013) under grant agreement no. 606862 (F-CHROMA), and from the Research Council of Norway through the Programme for Supercomputing. We are grateful to our anonymous referee whose insightful comments have improved this paper.


\section*{Data Availability}
 
The data and code used to obtain the results of this paper can be available at:
\href{https://github.com/miguelrq21-spec/Delay-observations-mid-infrared-channels-in-solar-flare-footpoints}{https://github.com/miguelrq21-spec}.\\
In this paper, we used the SunPy open-source software package \citep{sunpy_community2020} and the software library (\url{https://zenodo.org/records/14919949}) to analyse GOES data. All the figures within this paper were produced using Python colour-blind-friendly colour tables \citep[see][]{Wright17}. The GOES data is available at \url{https://www.ncei.noaa.gov/data/goes-space-environment-monitor/access/science/xrs/goes15/gxrs-l2-irrad_science/2014/09/} and analysis routines are available at \url{https://docs.sunpy.org/en/stable/generated/gallery/time_series/goes_xrs_example.html}.



\bibliographystyle{mnras}
\bibliography{example} 




\appendix
\label{appendix}
\section{Standard and Local Cross Correlation Function}
\label{A1}

The standard definition of the CCF of two time series $x_i$ and $y_i$ sampled at discrete times $t_i(i=1, \ldots, N)$ with equal sampling $\left(\Delta t=t_{i+1}-t_i\right)$ is

\begin{equation}
    CCF\left(\tau_k\right) \equiv \\ \qquad \frac{\frac{1}{N} \sum_{i=1}^{N-k}\left(x_i-\bar{x}\right)\left(y_{i+k}-\bar{y}\right)}{\left[\frac{1}{N} \sum_{i=1}^N\left(x_i-\bar{x}\right)^2\right]^{1 / 2}\left[\frac{1}{N} \sum_{i=1}^N\left(y_i-\bar{y}\right)^2\right]^{1 / 2}},   
    \label{eq:1}
\end{equation}

where the lag $\tau_k$ is the size of the time shift: $\tau_k=k \Delta t$, $k=0, \ldots, N-1$ and $\bar{x}, \bar{y}$ are the means of $x_i$ and $y_i$. \\

The suggested Local Cross Correlation function is defined as:

\begin{equation}
CCF_L\left(\tau_k\right)\equiv A \,
\end{equation}
where
\begin{equation}
A = \frac{\frac{1}{(N-k)} \sum_{i=1}^{N-k}\left(x_i-\bar{x}_*\right)\left(y_{i+k}-\overline{y_*}\right)}{\left[ \frac{1}{(N-k)} \sum_{i=1}^{N-k}\left(x_i-\bar{x}_*\right)^2\right]^{1 / 2}\left[\frac{1}{(N-k)} \sum_{i=k+1}^N\left(y_i-\bar{y}_*\right)^2\right]^{1 / 2}}
\end{equation}

where  $ \tau_k=k \Delta t $ and 

\begin{equation}
\overline{x_*}=\frac{1}{N-k} \sum_{i=1}^{N-k} x_i, \quad \overline{y_*}=\frac{1}{N-k} \sum_{i=k+1}^N y_i
\end{equation}


\section{ Comparison between standard and local cross correlation Function}
\label{A2}

\begin{figure*}
    \centering
    \includegraphics[width=1\linewidth]{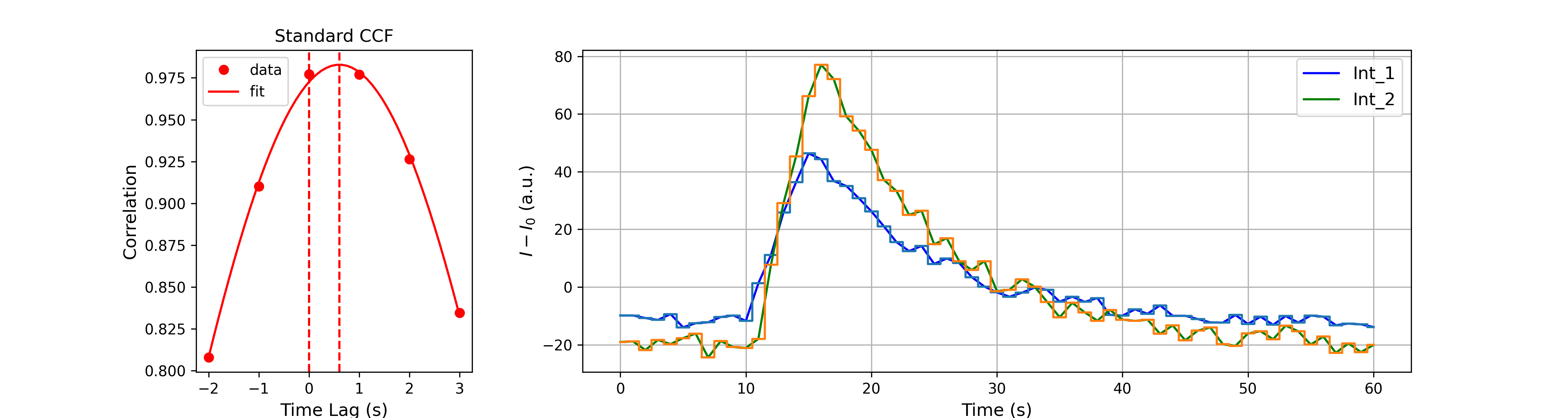}
    \caption{Two flare signals with identical profiles, but scaled in intensity, sampled each second but lagged relative to each other by a known value of \SI{0.75}{s}. Noise with an amplitude equal to 10 \% of the highest amplitude of the signal has been added. For lag estimation, the signal minus its mean is used. The estimated lag using the standard definition of the CCF is: \SI{0.67}{s}. The vertical red lines mark the maximum of the Gaussian function and the lag value equal to zero; the distance between the two then corresponds to the lag estimate.}
    \label{fig:1}
\end{figure*}

\begin{figure}
    \centering
    \includegraphics[width=1\linewidth]{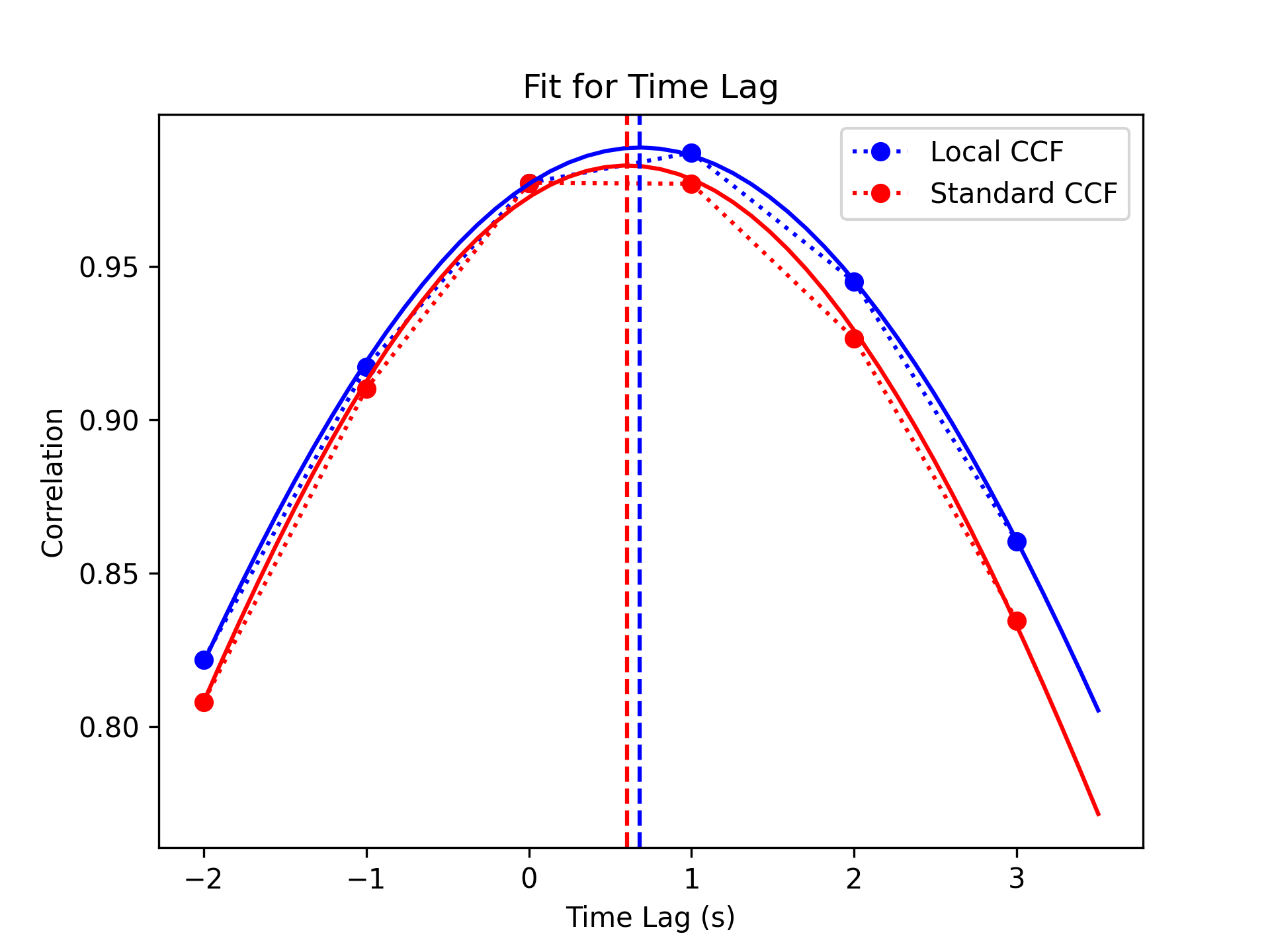}
    \caption{Differences in lag estimation from a well-known signal, using the standard definition of the CCF compared to using the local CCF. The Local CCF function provides a better estimate of \SI{0.74}{\second} with less bias towards zero. The standard definition of the CCF estimate a lag of \SI{0.67}{\second}.}
    \label{fig:3}
\end{figure}

To assess the performance of each definition, the local CCF was applied to a well-known noise-free signal lagged by \SI{0.75}{s} (see Fig.~\ref{fig:1}). This method yielded an estimated lag of \SI{0.74}{s}, whereas the standard CCF produced a value of \SI{0.69}{s}. The corresponding results are presented in Fig.~\ref{fig:3}. This comparison shows that, under ideal conditions (a known signal without noise), the standard correlation function method underestimates the lag between signals. A better estimate is obtained by using the local correlation function.

\bsp	
\label{lastpage}
\end{document}